\documentclass[aps,prl,twocolumn,showpacs,floatfix]{revtex4}
\usepackage{amsmath}
\usepackage{amsfonts}
\usepackage{amssymb}
\usepackage{graphics}
\begin{document}
\title{Protecting Bi-Partite Entanglement by Quantum Interferences}
\author{Sumanta Das$^{1,2}$ and G. S. Agarwal$^1$}
\affiliation{$^1$Department of Physics, Oklahoma State University, Stillwater,
Oklahoma 74078, USA\\
$^2$Institute for Quantum Science and Engineering and Department of Physics, Texas A$\&$M University, College Station, Texas 77843, USA}
\date{\today}
\begin{abstract}
We show that vacuum induced coherence in three level atomic systems can lead to preservation of bi-partite entanglement, when two such atoms are prepared as two initially entangled qubits each independently interacting with their respective vacuum reservoirs. We explicitly calculate the time evolution of concurrence for two different Bell states and show that a large amount of entanglement can survive in the long time limit. The amount of entanglement left among the two qubits depends strongly on the ratio of the non- orthogonal transitions in each qubit and can be more than $50\%$.
\end{abstract}
\maketitle
Successful implementation of many protocols in quantum information science requires sustained entanglement \cite{niel}. The effect of decoherence \cite{zurek} is unavoidable in systems like trapped ions/atoms \cite{blatt}, which are current-front runners in the practical realization of quantum information protocols. Thus devising ways or finding system where one can protect entanglement among quantum systems in the face of decoherence is a challenging question in the field of quantum information science. This has attracted immense interest in recent years and several methods have been proposed to avoid decoherence \cite{gsa01,zoller}.  Spontaneous emission - an intrinsic source of decoherence in many of these nanoscale systems needs to be inhibited to a large extent. For this purpose use of photonic band gap structure \cite{john94} and external fields \cite{gsa93} has been suggested. Note that all these proposals mainly concentrate on studying inhibition of decoherence in single qubits. Entanglement on the other hand involves two or more qubits and is known to decay much faster than the time scale over which single qubit coherences vanishes \cite{eberly}. For example, the typical decoherence time scale of a single qubit is $\tau_{s}\sim (\gamma)^{-1}$, where $2\gamma$ is the rate of spontaneous emission, whereas for two non-interacting entangled qubits it was found to be $\tau_{E} \sim (\gamma)^{-1} \ln [(2+\sqrt{2})/2]$ \cite{eberly}. Thus it is an open question whether some of the methods for manipulating decay in single qubit remain effective for entangled qubits. Some recent investigations have reported revival of two qubit entanglement for environments with memory as for example by using photonic band gap material \cite{belo} and by using coherent qubit-qubit interactions \cite{das}.\\
In an early work it was shown how quantum interference can lead to quenching of spontaneous emission in atomic systems \cite{gsab, scully1}. This comes about due to the interference between different channels of spontaneous emission and is known in the literature as vacuum induced coherence (VIC) \cite{gsab}. One of the key conditions for VIC to occur is that the dipole matrix elements of the two transitions from the excited states to the common ground state are nonorthogonal. Numerous theoretical studies thereafter have predicted fascinating applications of VIC \cite{ficek}. Recently VIC has been observed experimentally in emission from the trion states of quantum dots \cite{guru}. Rubidium atom is another system where effects of VIC were realized by fulfilling the conditions using external coherent drive \cite{wang}. Further it was reported that there is a variant of such vacuum mediated coupling with slightly relaxed conditions in the degenerate transitions $j = 1/2 \rightarrow j = 1/2$ of $Hg^{198+}$ and $Ba^{139+}$ ions \cite{kiff}.\\ 
Motivated by these recent developments we, in this paper study the effect of such quantum interference (VIC) on the entanglement dynamics of two initially entangled qubits formed out of two atomic systems in V-configuration. We show that the counter intuitive method of VIC can lead to preservation of bi-partite entanglement.  We in particular analyze to what extent entanglement initially present between two non-interacting qubits may be preserved.  We choose two qubits (bi-partite) for our study, as a well defined measure of entanglement for them, namely the concurrence \cite{wot} exist in the literature. We explicitly study the time evolution of concurrence and find that their exist a direct connection between the dynamics of entanglement and the single qubit excited state population.\\ 
\begin{figure}[h!]
\begin{center}
\scalebox{0.3}{\includegraphics{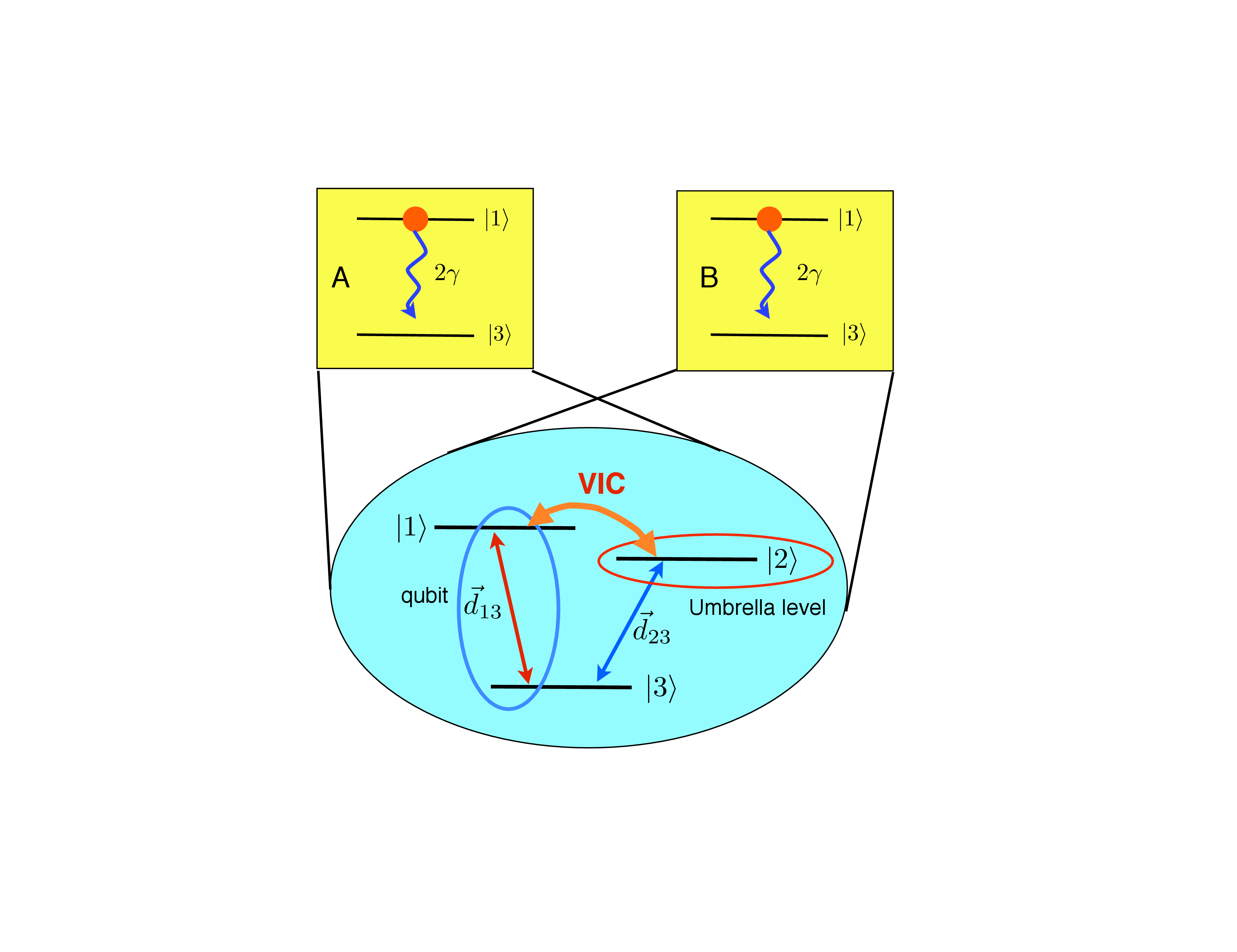}}
\caption{(Color online)Schematic diagram of two entangled non-interacting qubits A and B coupled to their independent environments. Each qubit can be envisaged by the levels $|1\rangle$ and $|3\rangle$ of a three level atomic system in a V- configuration.}
\end{center}
\end{figure}
Let us consider two qubits $A$ and $B$ (Fig. 1) formed by the excited and ground levels $|1\rangle$ and $|3\rangle$ of an atomic system. We assume that there exist no direct interaction among the qubits and that they are coupled independently to their respective vacuum reservoirs (environments). It has been shown that if these qubits are prepared initially in an entangled state then they become separable in a finite time given by $\tau_{E}$ even though decoherence of individual qubit takes infinite time \cite{eberly}. We next assume that their exits another level, $|2\rangle$ nearby to the excited level in each qubit (Fig 1). We name this, an \textit{umbrella level} for reasons that will be evident later. Note that the transition $|1\rangle \longrightarrow |2\rangle$ is considered to be forbidden in the electric dipole approximation. Now interaction of the system with the electromagnetic vacuum induces spontaneous decay of the two excited levels to the ground state. The two decay channels $|1\rangle \longrightarrow |3\rangle$ and $|2\rangle \longrightarrow |3\rangle$ leads to a quantum interference and gives rise to a cross-damping $2\gamma_{12} (1\leftrightarrow 2)$ among the two excited levels \cite{gsab}. The cross-damping term depends on the mutual orientation of the atomic transition dipole moments $\vec{d}_{13}$ and $\vec{d}_{23}$ and is given by $2\gamma_{12} \simeq \kappa [\vec{d}_{13}\cdot\vec{d}_{23}]$, where $\kappa$ is some constant \cite{gsab,ficek}. Note that there are several interesting consequences of this quantum interference \cite{ficek}, such as the coherent population trapping and generation of quantum coherences (VIC) in the excited states \cite{gsab,scully1}. We present in the following a detail analysis to show that such population trapping can considerably slow the environmental decoherence in entangled qubits thereby protecting entanglement.\\
We now present the mathematical model and calculation of time dependent measure of entanglement. We begin by first studying the dynamics of a single qubit formed out of the three-level system. The master equation that governs the dynamics of such a three level system interacting with the electromagnetic vacuum is given by \cite{gsab, ficek},
\begin{eqnarray}
\label{1}
\dot{\rho}& = & -\sum^{2}_{k = 1}\left\{i[\omega_{k}A_{kk}, \rho]+\gamma_{k}(A_{kk}\rho -2A_{3k}\rho A_{k3}+\rho A_{kk})\right\}\nonumber\\
& & -\gamma_{12}(A_{12}\rho -2A_{32}\rho A_{13}+\rho A_{12})+ (1\leftrightarrow 2),
\end{eqnarray}
where $\rho$ is the system density matrix, $A_{ij} = |i\rangle \langle j|$ are the atomic operators, $\omega_{1}$, $\omega_{2}$ are the transition frequencies and $2\gamma_{k}, (k = 1,2)$ are the decay rates for $|1\rangle \leftrightarrow |3\rangle$ and $|2\rangle \leftrightarrow |3\rangle$ respectively.  Note that for quantum interference to exits the cross-damping term $\gamma_{12} \neq 0$. This occurs if the dipole moments of the transition $|1\rangle\leftrightarrow |3\rangle$ and $|2\rangle\leftrightarrow |3\rangle$ are non-orthogonal, a condition usually difficult to meet. However it was later predicted that this stringent condition can be by-passed in an anisotropic vacuum \cite{gsa_prl}. Recently realization of VIC in presence of anisotropic vacuum has been proposed in left handed materials and metallic nanostructures \cite{zu}. Moreover some recent works with ions have also reported VIC effects in absence of the non-orthogonal transitions \cite{kiff}.
Since our qubit in this V-configuration is formed out of the two levels $|1\rangle$ and $|3\rangle$, we restrict our focus in studying the dynamics of this levels only. Now for simplicity we consider $\omega_{1} = \omega_{2}$ and $\gamma_{1} = \gamma, \gamma_{2} = \eta^{2}\gamma, \gamma_{12} = \gamma_{21} = \eta\gamma$. Here $\eta$ is a measure of the relative strengths of the decay on the transition $|1\rangle \leftrightarrow |3\rangle$ and $|2\rangle \leftrightarrow |3\rangle$ and is given by $\eta = |\vec{d}_{23}|/|\vec{d}_{13}|$, where we have assumed that the two transition moments are parallel.  On expanding eqn. (\ref{1}) in the basis defined by $|1\rangle$, $|2\rangle$, $|3\rangle$ and solving the coupled equations of population and coherences \cite{gsab} we get ,
\begin{eqnarray}
\label{2a}
\rho_{11}^{(s)}(t)& = & \frac{1}{2}\exp[-\gamma(1+\eta^{2})t](\rho_{11}^{(s)}(0)-\rho_{22}^{(s)}(0))\nonumber\\
&+&\frac{1}{2}\left[ \frac{2}{1+\eta^{2}}e^{-2\gamma(1+\eta^{2})t}-\frac{1-\eta^{2}}{1+\eta^{2}}e^{-\gamma(1+\eta^{2})t}\right]\alpha\nonumber\\
&+&\frac{1}{2}\left[ \frac{2\eta^{2}}{1+\eta^{2}}-\frac{1-\eta^{2}}{1+\eta^{2}}e^{-\gamma(1+\eta^{2})t}\right]\beta, \\
\rho_{33}^{(s)}(t)& = & 1-\exp[-2\gamma(1+\eta^{2})t]\alpha-\beta,\\
\rho_{13}^{(s)}(t)& = & \frac{1}{1+\eta^{2}}\left(\eta^{2}+\exp[-\gamma(1+\eta^{2})t] \right)\rho_{13}^{(s)}(0)\nonumber\\
& - & \frac{\eta}{1+\eta^{2}}\left(1-\exp[-\gamma(1+\eta^{2})t] \right)\rho_{23}^{(s)}(0),
\end{eqnarray} 
where $\alpha  = (1/2)(\rho_{11}^{(s)}(0)+\rho_{22}^{(s)}(0)+\rho_{12}^{(s)}(0)+\rho_{21}^{(s)}(0))$,
$\beta  = (1/2)(\rho_{11}^{(s)}(0)+\rho_{22}^{(s)}(0)-\rho_{12}^{(s)}(0)-\rho_{21}^{(s)}(0))$ and we have put the prefix $(s)$ to signify that the terms correspond to a single qubit. It can be clearly seen from the above solution that the population of the states $|1\rangle$ and $|3\rangle$ and thus the dynamics of the qubit is crucially dependent on population of state $|2\rangle$ and the coherences $\rho_{12}$ which arise due to VIC. In the steady state we find that the population of the excited state $|1\rangle$ is, $\rho_{11}^{(s)}(\infty) = [\eta^{2}/(1+\eta^{2})]\beta$ which is a non-zero constant for $\beta \neq 0$. Hence in presence of level $|2\rangle$ we find that a part of initial population can remain trapped in the state $|1\rangle$ in the long time limit thereby inhibiting spontaneous decay of the excited state. Level $|2\rangle$ thus act as a protection against environment induced decay of the qubit and hence we call it as the umbrella level. Note that in absence of level $|2\rangle$ the population of the state $|1\rangle$ decays as $e^{-2\gamma t}\rho_{11}(0)$ and thus vanishes as $t \rightarrow \infty$. \\
Till now we have been concerned about the dynamics involving only one qubit. We next consider a bi-partite system involving the qubits $A$ and $B$ together coupled to their respective vacuum reservoirs (environments). Note that in recent times such two qubit models have been extensively studied to investigate generation manipulation and dynamics of bi-partite entanglement and its response to system-environment coupling (decoherence). In the spirit of such studies and from the context of quantum information science we now ask an important question - can VIC contained decay of bi-partite (two qubit) entanglement ? \\
To investigate this we need to study the combined dynamics of the two qubits. We have already shown that as a result of vacuum induced coherences the decay of a single qubit gets inhibited. Whether and to what extent such coherences could influence an entangled two qubit system is a question we want to address now. Given the non-local character of entanglement, we would like to caution the educated reader at this point about making any pre-emptive assumption about the trivial extension of the single qubit to the two qubit cases. The assumption that the qubits are non-interacting and are independently coupled to their respective environment simplifies our analysis and yet can give valuable insight of the interplay of VIC and environment decoherence on the dynamics of entanglement. Most importantly this assumptions implies that the total time evolution operator for the two qubit system can be factorized into two terms corresponding to each qubit. As such the dynamics of each qubit governed by equation (\ref{1}) evolves independent of the other. \\
Let us now consider that at an initial time the two qubits are prepared in an entangled state given by one of the Bell states: $|\Psi\rangle = [|1_{A}1_{B}\rangle + |3_{A}3_{B}\rangle]/\sqrt{2}$. Then by following a procedure similar to \cite{belo} we can obtain the two-qubit reduced density matrix at time $t$ from the knowledge of single qubit dynamics. On doing the analysis we find the time evolved reduced density matrix $\rho^{(2)}(t)$ in the two qubit product basis $|1\rangle := |1_{A}1_{B}\rangle, |2\rangle := |1_{A}3_{B}\rangle, |3\rangle := |3_{A}1_{B}\rangle, |4\rangle := |3_{A}3_{B}\rangle$ to be,
 \begin{eqnarray}
 \label{3a} 
 \rho^{(2)}(t) = 
\left(\begin{array}{cccc} \rho^{(2)}_{11}(t) & 0 & 0 & \rho^{(2)}_{14}(t)\\
 0 & \rho^{(2)}_{22}(t) & 0 & 0\\
 0 & 0 & \rho^{(2)}_{33}(t) & 0 \\
\rho^{(2)}_{41}(t) & 0 & 0 & \rho^{(2)}_{44}(t)\ \end{array}\right),
\end{eqnarray}
where the prefix $(2)$ stands for the two qubit system and the matrix elements are given by,  
\begin{eqnarray}
\label{4a}
\rho^{(2)}_{11}(t) & = &\frac{1}{8(1+\eta^{2})} [\eta^{4} +e^{-4\gamma(1+\eta^{2})t}+2(1+\eta^{2})e^{-3\gamma(1+\eta^{2})t}\nonumber\\
&+&(1+\eta^{4}+4\eta^{2})e^{-2\gamma(1+\eta^{2})t}\nonumber\\
&+&2\eta^{2}(1+\eta^{2})e^{-\gamma(1+\eta^{2})t}],\nonumber\\
\rho^{(2)}_{22}(t) & = & \frac{1}{8(1+\eta^{2})} [\eta^{2} -e^{-4\gamma(1+\eta^{2})t}-(1+\eta^{2})e^{-3\gamma(1+\eta^{2})t}\nonumber\\
&+&(1-\eta^{2})e^{-2\gamma(1+\eta^{2})t}+(1+\eta^{2})e^{-\gamma(1+\eta^{2})t}],\nonumber\\
\rho^{(2)}_{22}(t) & = & \rho^{(2)}_{33}(t),\nonumber\\
\rho^{(2)}_{14}(t) & =& \frac{1}{2(1+\eta^{2})^{2}}\left [\eta^{2}+e^{-\gamma(1+\eta^{2})t}\right]^{2},
\end{eqnarray}
Note that $\rho^{(2)}(t)$ by itself is not normalized as it is projected out in a two dimensional space from the three dimensional space of the three level system in V-configuration. Since the existing measures for bi-partite entanglement requires the two qubit density matrix to be normalized, we normalize $\rho^{(2)}(t)$ by dividing it with $Tr(\rho^{(2)}(t))$.\\
To study the influence of VIC on the dynamics of bi-partite (two qubit) entanglement we next evaluate the two qubit concurrence $C(t)$ \cite{wot}. The concurrence, which is an entanglement monotone, is a well-defined and acceptable measure of bi-partite entanglement based on the two qubit density matrix. It varies from $0$ to $1$, where $C = 0$ is for separable qubits and $C = 1$ for maximally entangled qubits. Following the definition for concurrence \cite{wot} we find that for our system it is given by, 
\begin{equation}
\label{4}
C(t) = 2\mathsf{Max}\left \{ 0, |\rho^{(2)}_{14}(t)|-\sqrt{\rho^{(2)}_{22}(t)\rho^{(2)}_{33}(t)}\right\}.
\end{equation}
Note that for the Bell state $(|\Psi\rangle), C(0) = 1$. Thus initially the two qubits are maximally entangled. We see from eqn. (\ref{4}) that for the qubits to remain entangled at some later time $t$, concurrence $C(t) > 0$ and thus we need to satisfy the condition $\left [|\rho^{(2)}_{14}(t)| / \sqrt{\rho^{(2)}_{22}(t)\rho^{(2)}_{33}(t)}\right ]-1 > 0$. 
\begin{figure}[h!]
\begin{center}
\scalebox{0.29}{\includegraphics{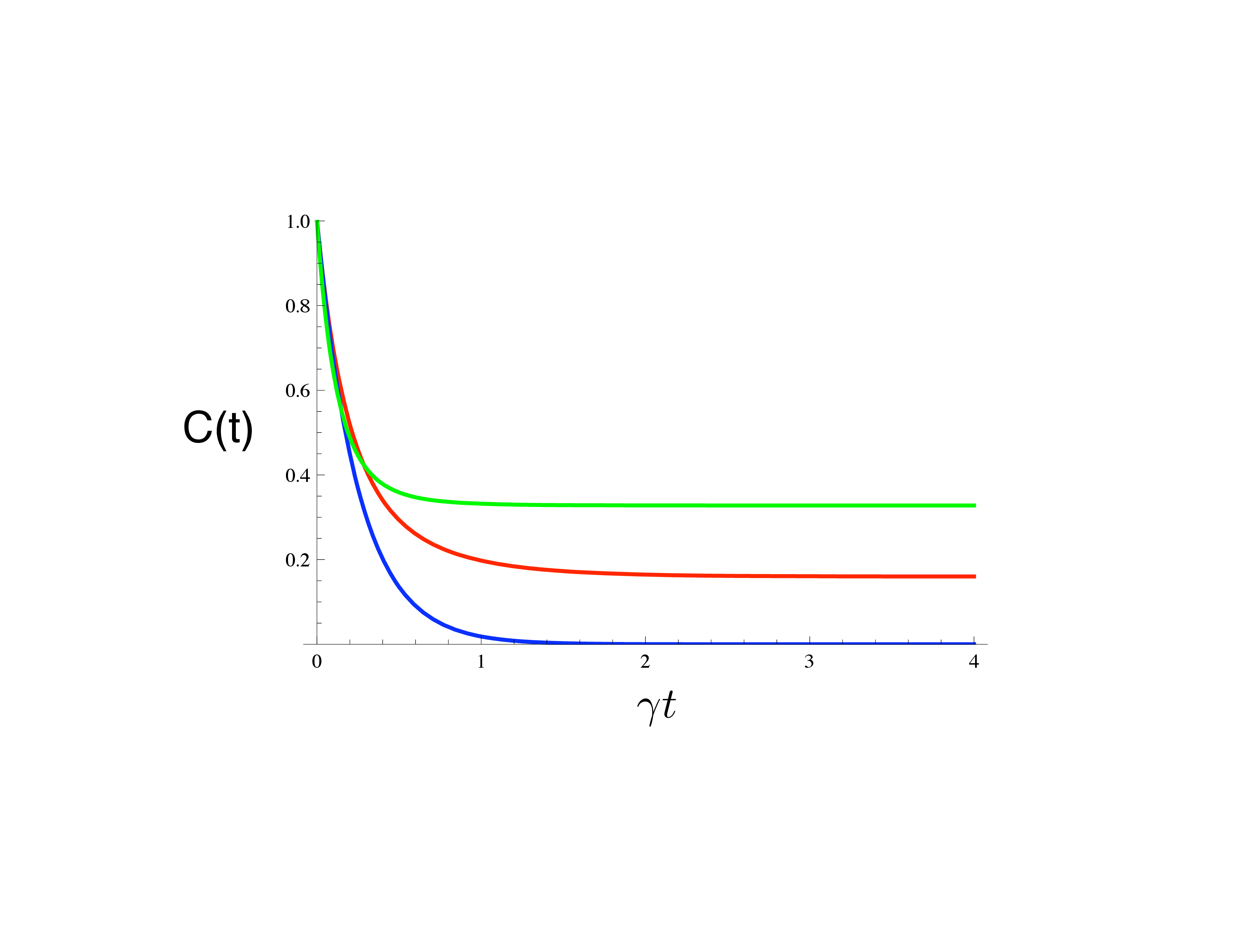}}
\caption{(Color online)Concurrence in presence and absence (blue curve) of VIC for the qubits prepared initially in the entangled state $|\Psi\rangle$. The red and green curves are for $\eta = 1, \sqrt{2}$ respectively. Here $\eta = \sqrt{2}$ implies that level $|2\rangle$ decays faster than the qubit. Substantial entanglement remains among the qubits even after a long time due to VIC effects.}
\end{center}
\end{figure}
In Fig. (2) we plot the time dependent concurrence of the Bell state $|\Psi\rangle$ as a function of normalized time $\gamma t$ for two different values of $\eta$. We see from Fig. (2), that in presence of the umbrella level $|2\rangle$, VIC effects lead to a slower decay of the concurrence with increasing time. Further in the long time limit and in presence of VIC, concurrence becomes time independent and substantial amount of initial entanglement is left among the qubits, whereas in absence of VIC, entanglement vanishes. Moreover, we find that the amount of entanglement left is sensitive to the value of $\eta$ and for higher values of $\eta$ the rate of decay becomes much slower at a later time. We find that for $\eta = \sqrt{2}$, almost $40\%$ of the initial entanglement is left. Remember that $\eta$ was defined as the relative strengths of the decay along the two transitions $|1\rangle \leftrightarrow |3\rangle$ and $|2\rangle \leftrightarrow |3\rangle$ and thereby is related to the rate of decay of level $|2\rangle$ in comparison to decay rate of the qubit. The behavior of concurrence as seen from Fig. (2) thus implies that, faster the decay of the umbrella level with respect to the qubit, more effective is VIC in countering the environmental decoherence. The dynamics of entanglement can be understood by studying the analytical expressions (\ref{4a}) and (\ref{4}). In the limit ($t = \infty$), we find that $\rho^{(2)}_{14}(\infty)/\sqrt{\rho^{(2)}_{22}(\infty)\rho^{(2)}_{33}(\infty)} = [4\eta^{2}/(1+\eta^{2})] > 1$, and thus $C(\infty)$ is a non-zero constant that depends on $\eta$ only. Moreover from eqn. (\ref{4a}) we find that, $\rho^{(2)}_{11}(\infty) = [(\eta^{4}/8)/(1+\eta^{2})]$ in comparison to $[\rho^{(2)}_{11}(t) = e^{-4\gamma t} \rho^{(2)}_{11}(0) \rightarrow \rho^{(2)}(\infty) = 0]$ when there is no VIC. Thus we see that, VIC leads to population trapping in the doubly excited state and serves as a protector of entanglement against environmental decoherence in the two qubit system.\\  
We next consider the two qubits initially prepared in the other Bell state, $|\Phi\rangle = [|1_{A}3_{B}\rangle + |3_{A}1_{B}\rangle]/\sqrt{2}$ which is maximally entangled at $t = 0$. Following the above analysis we find the time dependent concurrence in this case to be: $C(t) = 2\mathsf{Max}\{ 0, |\rho^{(2)}_{23}(t)|\}$, where $\rho^{(2)}_{23}(t)$ is given by, 
\begin{eqnarray}
\label{6a}
\rho^{(2)}_{23}(t)  =  \frac{1}{2 Tr\{\rho^{(2)}(t)\}(1+\eta^{2})^{2}}\left [\eta^{2}+e^{-\gamma(1+\eta^{2})t}\right]^{2}.
\end{eqnarray}
\begin{figure}[h!]
\begin{center}
\scalebox{0.295}{\includegraphics{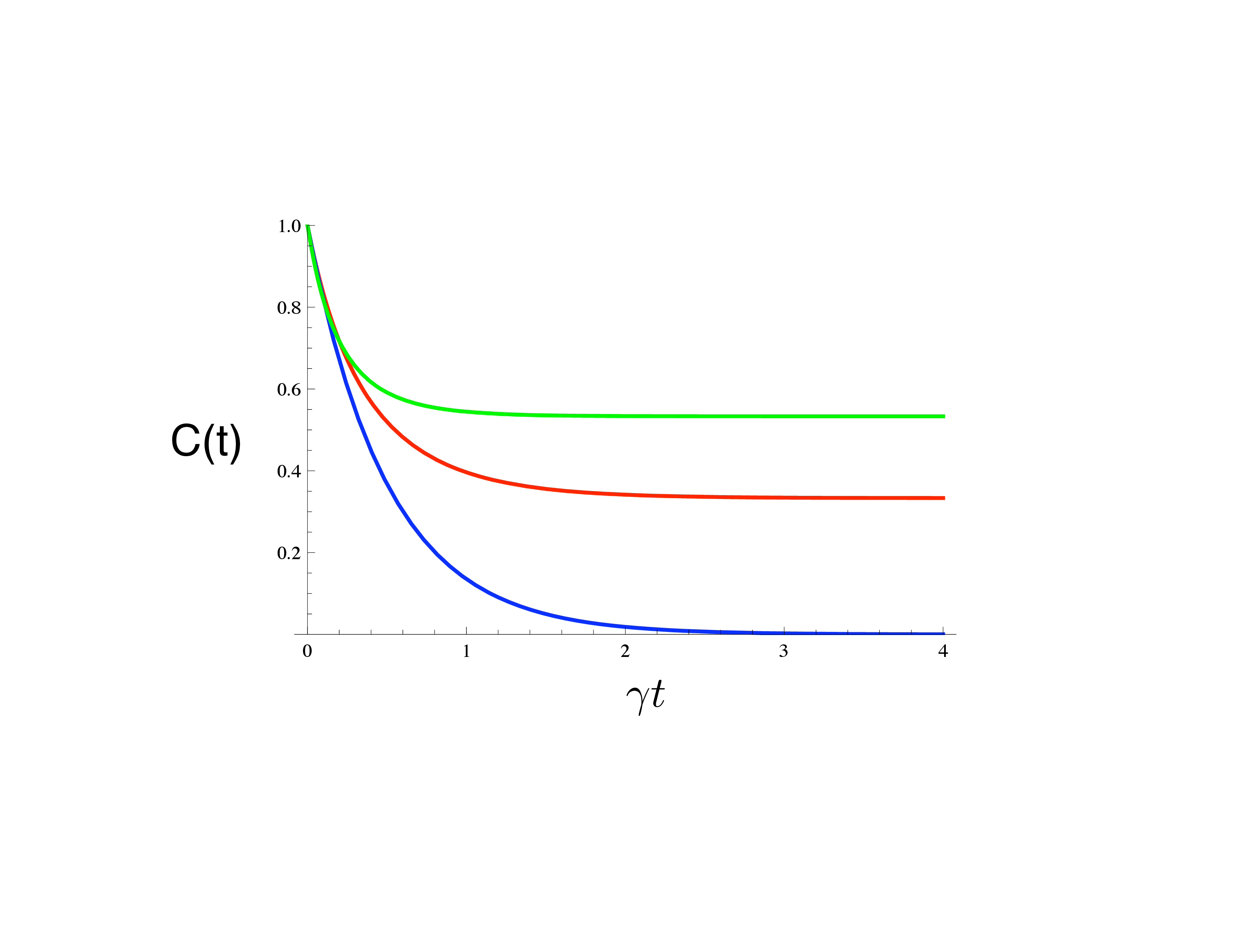}}
\caption{(Color online)Concurrence in presence and absence (blue curve) of VIC for the qubits prepared initially in the entangled state  $|\Phi\rangle$. The red and green curves are for $\eta = 1, \sqrt{2}$ respectively. Almost $60\%$ of the initial entanglement among the qubits can be preserved due to VIC.}
\end{center}
\end{figure}
In Fig. (3) we plot the dynamical evolution of entanglement for the state $|\Phi\rangle$ as a function of normalized time $\gamma t$ for two different values of $\eta$. The behavior of concurrence $C(t)$ is found to be similar to that of Fig. (2). We see that for the state $|\Phi\rangle$ in the long time limit and in presence of VIC for $\eta = \sqrt{2}$, almost $60\%$ of the initial entanglement among the qubits remains. The time independence of the concurrence as $t \rightarrow \infty$ can be understood from the expression of $C(t)$ and (\ref{6a}). We find that $\rho^{(2)}_{23}(\infty)  = \eta^{4}/[2Tr\{\rho^{(2)}(\infty)\}(1+\eta^{2})^2] > 0$ and thus $C(\infty)$ is a non-zero constant which depend on $\eta$.\\
Our analysis thus convincingly show that VIC can substantially control spontaneous decay in entangled two qubit systems thereby protecting entanglement to a large extent against environmental decoherence.

\end{document}